\RequirePackage{fix-cm} 
\documentclass[a4paper, oneside, reqno, dvips, 11pt]{extarticle}
\usepackage{fixltx2e}   

\usepackage[latin1]{inputenc}
\usepackage[T1]{fontenc}

\usepackage{eucal}
\usepackage{xspace}
\usepackage{amsgen}
\usepackage{amsmath}
\usepackage{MnSymbol}
\usepackage{mathtools}
\usepackage[nice]{nicefrac}
\usepackage{mathrsfs, dsfont}
\usepackage{esint}

\usepackage{bm}

\usepackage{a4wide}

\usepackage[table]{xcolor}
\usepackage{dcolumn}  

\definecolor{oneblue}{rgb}{0.0, 0.0, 0.85}

\usepackage{psfrag}
\usepackage{graphicx}
\usepackage{subfigure}
\usepackage{morefloats}
\usepackage{indentfirst}

\usepackage{acronym}
\usepackage{microtype}
\usepackage[labelfont={bf,sf},%
            labelsep=period,%
            justification=raggedright]{caption}

\usepackage[usenames, dvipsnames, pdf]{pstricks}
\usepackage{epsfig}
\usepackage{pst-grad} 
\usepackage{pst-plot} 

\usepackage[colorlinks,
            urlcolor=oneblue,
            linkcolor=oneblue,
            citecolor=oneblue,
            bookmarksopen=false]{hyperref}
\hypersetup{pdfpagemode=UseNone}

\usepackage{titletoc}
\usepackage{titlesec}
\titleformat{\section}{\large\bfseries\scshape}{\thesection.}{0.5em}{}

\numberwithin{equation}{section}

\newcommand{\ie}{\emph{i.e.}~}
\newcommand{\eg}{\emph{e.g.}~}

\title{\bf Modeling water waves beyond perturbations\bigskip}
\author{
        Didier \textsc{Clamond} \\
        {\small Universit\'e de Nice -- Sophia Antipolis} \\
        {\small Laboratoire J.~A. Dieudonn\'e} \\
        {\small Parc Valrose, 06108 Nice cedex 2, France} \and
        Denys \textsc{Dutykh} \\
        {\small CNRS--LAMA UMR 5127} \\
        {\small Universit\'e Savoie Mont Blanc} \\
        {\small 73376 Le Bourget-du-Lac France}}

\date{\today}

\begin{document}

\maketitle

\begin{abstract}
In this chapter, we illustrate the advantage of variational principles for modeling water waves from an elementary practical viewpoint. The method is based on a `relaxed' variational principle, \ie, on a Lagrangian involving as many variables as possible, and imposing some suitable subordinate constraints. This approach allows the construction of approximations without necessarily relying on a small parameter. This is illustrated via simple examples, namely the Serre equations in shallow water, a generalization of the Klein--Gordon equation in deep water and how to unify these equations in arbitrary depth. The chapter ends with a discussion and caution on how this approach should be used in practice.
\end{abstract}

\tableofcontents
\clearpage

\section{Introduction}

Surface water waves are a very rich physical phenomenon with a long research history \cite{Craik2004,Wehausen1960}. In addition to their fundamental physical importance, understanding water waves is also important for many applications related to human safety and economy such as tsunamis, freak waves, harbor protections, beach nourishment/erosion, just to mention a few examples.  Water waves are a paradigm for many nonlinear wave phenomena in various physical media. The prominent physicist Richard P. Feynman wrote in his cerebrated lectures \cite{Feynman1964}: 
{\em ``Water waves that are easily seen by everyone, and which are usually used as an example of waves in elementary courses, are the worst possible example; they have all the complications that waves can have.''}
This is precisely these complications that make the richness and interest of water waves. Indeed, despite numerous studies, new waves and new wave behaviors are still discovered (\eg, \cite{Rajchenbach2013, Rajchenbach2011}) and wave dynamics is still far from being fully understood.

Mathematical and numerical models are unavoidable for understanding water waves. Although the primitive equations governing these waves are rather simple to write, their mathematical analysis is highly non trivial and even their numerical resolution is very demanding. Therefore, simplified models are crucial to gain insight and to derive operational numerical models. Most of the time, simplified models are derived via some asymptotic expansions, exploiting a small parameter in the problem at hands. This approach is very effective leading to well-known equations, such as the Saint-Venant \cite{Stoker1957, Wehausen1960}, Boussinesq \cite{bouss}, Serre--Green--Naghdi \cite{Green1976, Serre1953}, Korteweg-deVries \cite{KdV} equations in shallow water and the nonlinear Schr\"odinger \cite{Mei1989}, Dysthe \cite{Dysthe1979} equations in deep water.  These equations being most often derived via some perturbation techniques, they are valid for waves of small amplitude or/and small {\em wavelength\//\/water depth\/} ratio. However, for many applications it is necessary to use models uniformly valid for all depths and that are accurate for large amplitudes. Moreover, some phenomena \cite{Rajchenbach2013, Rajchenbach2011} do not involve any small parameter and do not bifurcate from rest. The problem is then to derive models without relying on a small parameter.

It is well-known in theoretical physics that variational formulations are tools of choice to derive approximations when small parameter expansions are inefficient. Fortunately, a variational principle is available for water waves that can be exploited to derive approximations. There are mainly two variational formulations for irrotational surface waves that are commonly used, namely the Lagrangian of Luke \cite{Luke1967} and the Hamiltonian of Broer, Petrov and Zakharov \cite{Broer1974, Petrov1964, Zakharov1968}. Details on the variational formulations for surface waves can be found in review papers, \eg, \cite{Radder1999, Salmon1988, Zakharov1997}.

In water wave theory, variational formulations are generally used together with a small parameter expansion. This is not necessary, however, because variational methods can also be fruitfully used without small parameter, as it is well-known in Quantum Mechanics, for example. This was demonstrated in \cite{Clamond2009}, the present chapter being a simpler illustration of this idea, with some complementary remarks. A companion presentation with further comments can be found in \cite{Clamond2014}. Here, only elementary knowledge in vector calculus is assumed, as well as some familiarity with the Euler--Lagrange equations and variational principles in Mechanics \cite{Goldstein2001,Lanczos1970}.

The Chapter is organized as follows. In section~\ref{sec:model}, the physical hypothesis, notations and equations are given for the classical problem of irrotational surface gravity waves. In section~\ref{sec:lag}, Luke's Lagrangian is {\em relaxed\/} to incorporate explicitly more degrees of freedom. This modification yields the Hamilton principle in its most general form. The advantage of this formulation is subsequently illustrated with examples over a fixed horizontal bottom, for the sake of simplicity. We begin with a shallow water model, followed by a deep water one and ending with an arbitrary depth generalization. Further generalizations, shortcomings and perspectives are discussed in section~\ref{secdis}.

\section{Preliminaries}\label{sec:model}

\begin{figure}
\centering
\scalebox{0.75} 
{
\begin{pspicture}(0,-3.545)(16.039062,3.454)
\definecolor{color32}{rgb}{0.12156862745098039,0.08627450980392157,0.8352941176470589}
\definecolor{color96}{rgb}{0.3254901960784314,0.1607843137254902,0.1607843137254902}
\definecolor{color145}{rgb}{0.054901960784313725,0.01568627450980392,0.01568627450980392}
\definecolor{color145b}{rgb}{0.6549019607843137,0.44313725490196076,0.44313725490196076}
\definecolor{color168}{rgb}{0.09411764705882353,0.0392156862745098,0.0392156862745098}
\definecolor{color168b}{rgb}{0.06274509803921569,0.023529411764705882,0.023529411764705882}
\definecolor{color423}{rgb}{0.2549019607843137,0.27058823529411763,0.9372549019607843}
\definecolor{color558}{rgb}{0.22745098039215686,0.11372549019607843,0.11372549019607843}
\psline[linewidth=0.028cm,linestyle=dashed,arrowsize=0.1cm 2.0,arrowlength=1.4,
arrowinset=0.4]{->}(0.0,1.68)(16.0,1.68)
\psline[linewidth=0.028cm,linestyle=dashed,arrowsize=0.1cm 2.0,arrowlength=1.4,
arrowinset=0.4]{<-}(8.0,2.9)(8.0,-3.1)
\pscustom[linewidth=0.044,linecolor=color32]
{
\newpath
\moveto(0.96,1.68)
\lineto(1.2286792,1.68)
\curveto(1.3630188,1.68)(1.732453,1.8)(1.9675473,1.92)
\curveto(2.2026415,2.04)(2.67283,2.04)(2.9079244,1.92)
\curveto(3.1430187,1.8)(3.747547,1.52)(4.116981,1.36)
\curveto(4.486415,1.2)(5.359623,1.2)(5.863396,1.36)
\curveto(6.36717,1.52)(7.072453,1.8)(7.2739625,1.92)
\curveto(7.475472,2.04)(7.8784904,2.12)(8.08,2.08)
\curveto(8.281509,2.04)(8.718113,1.84)(8.953207,1.68)
\curveto(9.188302,1.52)(9.591321,1.32)(9.759245,1.28)
\curveto(9.927169,1.24)(10.263018,1.32)(10.430943,1.44)
\curveto(10.598867,1.56)(10.901132,1.76)(11.035471,1.84)
\curveto(11.169811,1.92)(11.472075,2.0)(11.64,2.0)
\curveto(11.807925,2.0)(12.177359,2.0)(12.378868,2.0)
\curveto(12.580379,2.0)(12.949812,1.96)(13.117737,1.92)
\curveto(13.285662,1.88)(13.621509,1.8)(13.7894335,1.76)
\curveto(13.957358,1.72)(14.293208,1.68)(14.461133,1.68)
\curveto(14.629058,1.68)(14.897737,1.68)(15.2,1.68)
}
\psline[linewidth=0.05cm,linecolor=color96](0.32,-2.64)(1.6281435,-2.64)
\psline[linewidth=0.05cm,linecolor=color96](14.055508,-2.64)(15.494465,-2.64)
\pscustom[linewidth=0.05,linecolor=color96]
{
\newpath
\moveto(1.6281437,-2.64)
\lineto(1.8897723,-2.56)
\curveto(2.0205867,-2.52)(2.2495117,-2.4)(2.3476226,-2.32)
\curveto(2.4457333,-2.24)(2.609251,-2.28)(2.6746583,-2.4)
\curveto(2.7400653,-2.52)(2.8708801,-2.8)(2.9362872,-2.96)
\curveto(3.0016944,-3.12)(3.2306194,-3.24)(3.3941376,-3.2)
\curveto(3.5576556,-3.16)(3.7865806,-2.96)(3.8519876,-2.8)
\curveto(3.9173946,-2.64)(4.0809126,-2.36)(4.1790233,-2.24)
\curveto(4.2771344,-2.12)(4.5060596,-2.12)(4.6368737,-2.24)
\curveto(4.767688,-2.36)(4.8985023,-2.68)(4.8985023,-2.88)
\curveto(4.8985023,-3.08)(4.996613,-3.36)(5.094724,-3.44)
\curveto(5.1928344,-3.52)(5.42176,-3.44)(5.5525746,-3.28)
\curveto(5.6833887,-3.12)(5.912314,-2.76)(6.0104246,-2.56)
\curveto(6.1085353,-2.36)(6.337461,-2.2)(6.468275,-2.24)
\curveto(6.599089,-2.28)(6.7953105,-2.48)(6.860718,-2.64)
\curveto(6.926125,-2.8)(7.1550503,-3.0)(7.318568,-3.04)
\curveto(7.482086,-3.08)(7.711011,-3.0)(7.7764187,-2.88)
\curveto(7.8418255,-2.76)(8.038047,-2.52)(8.168861,-2.4)
\curveto(8.299675,-2.28)(8.528601,-2.28)(8.626712,-2.4)
\curveto(8.724822,-2.52)(8.88834,-2.8)(8.953748,-2.96)
\curveto(9.019155,-3.12)(9.24808,-3.2)(9.411598,-3.12)
\curveto(9.575116,-3.04)(9.804041,-2.8)(9.869449,-2.64)
\curveto(9.934856,-2.48)(10.098374,-2.24)(10.196485,-2.16)
\curveto(10.294595,-2.08)(10.490815,-2.12)(10.588927,-2.24)
\curveto(10.687038,-2.36)(10.817853,-2.64)(10.850556,-2.8)
\curveto(10.883259,-2.96)(11.014074,-3.12)(11.1121855,-3.12)
\curveto(11.210296,-3.12)(11.373814,-2.96)(11.43922,-2.8)
\curveto(11.504626,-2.64)(11.668144,-2.36)(11.766256,-2.24)
\curveto(11.8643675,-2.12)(12.027885,-2.04)(12.093292,-2.08)
\curveto(12.158699,-2.12)(12.289515,-2.28)(12.35492,-2.4)
\curveto(12.420327,-2.52)(12.583845,-2.68)(12.681957,-2.72)
\curveto(12.780068,-2.76)(13.008993,-2.84)(13.139807,-2.88)
\curveto(13.270621,-2.92)(13.499546,-2.88)(13.597656,-2.8)
\curveto(13.695766,-2.72)(13.859284,-2.64)(14.055508,-2.64)
}
\usefont{T1}{ptm}{m}{n}
\rput(15.664532,1.31){$\bm{x}$}
\usefont{T1}{ptm}{m}{n}
\rput(7.7,2.8){$y$}
\usefont{T1}{ptm}{m}{n}
\rput(7.534531,1.31){$O$}
\psline[linewidth=0.03cm,linestyle=dotted,linecolor=color145,fillcolor=color145b,
arrowsize=0.1cm 2.0,arrowlength=1.4,arrowinset=0.4]{<->}(11.2,1.6)(11.2,-3.0)
\usefont{T1}{ptm}{m}{n}
\rput(10.444531,-0.45){$d(\bm{x},t)$}
\psline[linewidth=0.03cm,linestyle=dotted,linecolor=color168,fillcolor=color168b,
arrowsize=0.1cm 2.0,arrowlength=1.4,arrowinset=0.4]{<->}(4.48,1.2)(4.48,-2.1)
\usefont{T1}{ptm}{m}{n}
\rput(3.7245312,-0.45){$h(\bm{x},t)$}
\psline[linewidth=0.03cm,linestyle=dotted,linecolor=color168,fillcolor=color168b,
arrowsize=0.1cm 2.0,arrowlength=1.4,arrowinset=0.4]{>-<}(2.4,1.45)(2.4,2.25)
\usefont{T1}{ptm}{m}{n}
\rput(3.1,2.3){$\eta(\bm{x},t)$}
\psline[linewidth=0.05cm,linecolor=color423,fillcolor=color168b](0.8,0.4)(1.12,0.4)
\psline[linewidth=0.05cm,linecolor=color423,fillcolor=color168b](0.8,-1.68)(1.12,-1.68)
\psline[linewidth=0.05cm,linecolor=color423,fillcolor=color168b](2.72,0.72)(3.04,0.72)
\psline[linewidth=0.05cm,linecolor=color423,fillcolor=color168b](2.72,-1.52)(3.04,-1.52)
\psline[linewidth=0.05cm,linecolor=color423,fillcolor=color168b](5.6,0.56)(5.92,0.56)
\psline[linewidth=0.05cm,linecolor=color423,fillcolor=color168b](1.6,-0.4)(1.92,-0.4)
\psline[linewidth=0.05cm,linecolor=color423,fillcolor=color168b](5.92,-1.68)(6.24,-1.68)
\psline[linewidth=0.05cm,linecolor=color423,fillcolor=color168b](7.04,0.24)(7.36,0.24)
\psline[linewidth=0.05cm,linecolor=color423,fillcolor=color168b](9.6,0.56)(9.92,0.56)
\psline[linewidth=0.05cm,linecolor=color423,fillcolor=color168b](8.96,-1.68)(9.28,-1.68)
\psline[linewidth=0.05cm,linecolor=color423,fillcolor=color168b](12.8,0.72)(13.12,0.72)
\psline[linewidth=0.05cm,linecolor=color423,fillcolor=color168b](14.24,-1.68)(14.56,-1.68)
\psline[linewidth=0.05cm,linecolor=color423,fillcolor=color168b](13.12,-0.56)(13.44,-0.56)
\psline[linewidth=0.05cm,linecolor=color423,fillcolor=color168b](11.84,-1.52)(12.16,-1.52)
\psline[linewidth=0.05cm,linecolor=color423,fillcolor=color168b](8.8,-0.4)(9.12,-0.4)
\psline[linewidth=0.03cm,linecolor=color558,fillcolor=color168b](0.8,-2.64)(0.48,-2.96)
\psline[linewidth=0.03cm,linecolor=color558,fillcolor=color168b](1.28,-2.64)(0.96,-2.96)
\psline[linewidth=0.03cm,linecolor=color558,fillcolor=color168b](1.76,-2.64)(1.44,-2.96)
\psline[linewidth=0.03cm,linecolor=color558,fillcolor=color168b](14.24,-2.64)(13.92,-2.96)
\psline[linewidth=0.03cm,linecolor=color558,fillcolor=color168b](14.72,-2.64)(14.4,-2.96)
\psline[linewidth=0.03cm,linecolor=color558,fillcolor=color168b](15.2,-2.64)(14.88,-2.96)
\end{pspicture} 
}
\caption{\small\em Definition sketch.}
\label{fig:sketch}
\end{figure}
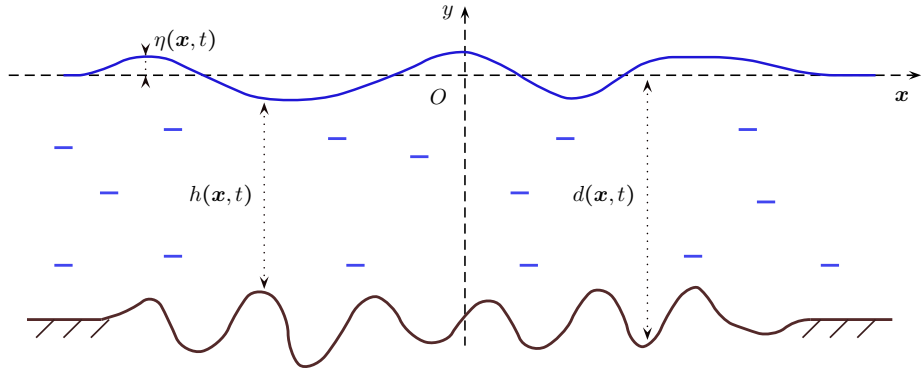

Consider an ideal incompressible fluid of constant density $\rho$. The horizontal independent variables are denoted by $\bm{x}=(x_1, x_2)$ and the upward vertical one by $y$. The origin of the Cartesian coordinate system is chosen such that the surface $y=0$ corresponds to the still water level. The fluid is bounded below by the bottom at $y=-d(\bm{x},t)$ and above by the free surface at $y = \eta(\bm{x},t)$. Usually, we assume that the total depth $h(\bm{x},t)\equiv d(\bm{x},t) + \eta(\bm{x},t)$ remains positive $h (\bm{x},t)\geqslant h_0 > 0$ at all times $t$ for some constant $h_0$. A sketch of the physical domain is shown on Figure~\ref{fig:sketch}.

We denote $\bm{u}=(u_1,u_2)$ the horizontal velocity and $v$ vertical one. The fluid density being constant, the mass conservation implies an isochoric motion yielding the continuity equation valid everywhere in the fluid domain
\begin{equation}\label{eqincexa}
  \bm{\nabla\cdot u}\ +\ \partial_y\/v\ =\ 0,
\end{equation}
where $\bm{\nabla}$ denotes the horizontal gradient and $\bm{\cdot}$ denotes the scalar (inner) product of vectors.

Denoting with over `tildes' and `breves' the quantities computed, respectively, at the free surface $y = \eta(\bm{x},t)$ and at the bottom $y=-d(\bm{x},t)$, the impermeabilities of these boundaries give the relations
\begin{align} \label{eqimpexa}
  \partial_t\/\eta\ +\ \tilde{\bm u}\bm{\cdot}\bm{\nabla}\eta\ =\ \tilde{v}, \qquad
  \partial_t\/d\ +\ \breve{\bm u}\bm{\cdot}\bm{\nabla} d\ =\ -\/\breve{v}.
\end{align}

Traditionally in water wave modeling, the assumption of flow irrotationality is also adopted because it is relevant in many situations and it brings considerable simplifications. The zero-curl velocity field condition can be written
\begin{equation}\label{eqirrexa}
  \bm{\nabla}\/v\ =\ \partial_y\/\bm{u}, \qquad 
  \bm{\nabla\times u}\ =\ 0,
\end{equation}
where ${\bm \times}$ is  a two dimensional analog of the cross product.\footnote{For two two-dimensional vectors ${\bm a}=(a_1,a_2)$ and ${\bm b}=(b_1,b_2)$, $\bm{a\times b}=a_1b_2 - a_2b_1$ is a scalar.} The irrotationality conditions (\ref{eqirrexa}) are satisfied identically introducing a (scalar) velocity potential $\phi$ such that
\begin{equation}\label{uvphi}
  {\bm u}\ =\ {\bm \nabla}\/\phi, \qquad v\ =\ \partial_y\/\phi.
\end{equation}
For irrotational motions of incompressible fluids, the Euler momentum equations can be integrated into the scalar Lagrange--Cauchy equation 
\begin{equation}
  p\ +\ \partial_t\/\phi\ +\ g\/y\ +\ {\textstyle{1\over2}}|\bm{\nabla}\phi|^2\ +\ {\textstyle{1\over2}}(\partial_y\phi)^2\ =\ 0,
\end{equation}
where $p$ is the pressure divided by the density $\rho$ and $g>0$ is the acceleration due to gravity. At the free surface the pressure is zero --- \ie, $\tilde{p}=0$ --- but surface tensions or other effects could be taken into account. Note that for steady flows, \ie when the velocity field is independent of time, $\partial_t\/\phi=\mathrm{constant}=-B$ and the Lagrange--Cauchy equation becomes the Bernoulli equation, $B$ being a Bernoulli constant.

In summary, with the hypotheses above, the governing equations of the classical (non overturning) surface water waves are \cite{Johnson2004, Stoker1958, Whitham1999}:
\begin{align}
  \bm{\nabla}^2\phi\ +\ \partial_y^{\,2}\phi\ =&\ 0, 
  \qquad -d(\bm{x},t)\leqslant y\leqslant\eta(\bm{x},t), \label{eq:laplace} \\
  \partial_t\eta\ +\ (\bm{\nabla}\phi)\bm{\cdot}(\bm{\nabla}\eta)\ -\ \partial_y\phi\ =&\ 0, 
  \qquad y = \eta(\bm{x}, t), \label{eq:kinematic} \\
  \partial_t\phi\ +\ {\textstyle{1\over2}}|\bm{\nabla}\phi|^2\ +\ {\textstyle{1\over2}}(\partial_y\phi)^2\ 
  +\ g\eta\ =&\ 0, 
  \qquad y = \eta(\bm{x},t), \label{eq:bernoulli} \\
  \partial_t d\ +\ (\bm{\nabla} d)\bm{\cdot}(\bm{\nabla}\phi)\ +\ \partial_y\phi\ =&\ 0, 
  \qquad y = -d(\bm{x},t). \label{eq:bottomkin}
\end{align}

The assumptions of fluid incompressibility and flow irrotationality lead to the Laplace equation \eqref{eq:laplace} for the velocity potential $\phi(\bm{x},y,t)$. The main difficulty of the water wave problem lies on the boundary conditions. Equations \eqref{eq:kinematic} and \eqref{eq:bottomkin} express the free-surface kinematic condition and bottom impermeability, respectively, while the dynamic condition \eqref{eq:bernoulli} expresses the free surface isobarity.

\section{Variational formulations} \label{sec:lag}

Equations (\ref{eq:laplace})--(\ref{eq:bottomkin}) can be derived from the ``stationary point'' (point where the variation is zero) of the following functional 
\[
  \mathcal{L}\ =\ \int_{t_1}^{t_2}\!\int_{\Omega}\mathscr{L}\,\rho\,\mathrm{d}^2\,{\bm x}\,\mathrm{d}t
\] 
($\Omega$ the horizontal domain) where the Lagrangian density $\mathscr{L}$ is \cite{Luke1967}
\begin{equation}\label{defL0}
  \mathscr{L}\,=\,-\int_{-d}^\eta\left[\,g\/y\,+\,\partial_t\/\phi\,+\,{\textstyle{1\over2}}\/|{\bm \nabla}\/\phi|^2\,+\,{\textstyle{1\over2}}\/(\partial_y\/\phi)^2\,\right]\mathrm{d}\/y.
\end{equation}
One can check that the Euler--Lagrange equations for this functional yield directly the water wave equations. (Detailed algebra can be found in \cite{Luke1967}, but also on Wikipedia.\footnote{{\sf http://en.wikipedia.org/wiki/Luke's\_variational\_principle}.})

Integrating by parts and neglecting the terms at the horizontal and temporal boundaries because they do not contribute to the functional variations (this will be done repeatedly below without explicit mention), Luke's variational formulation (\ref{defL0}) can be rewritten with the following Lagrangian density:
\begin{equation}\label{defL1}
  \mathscr{L}\ =\ \tilde{\phi}\,\eta_t\ +\ \breve{\phi}\,d_t\ -\ \frac{g\,\eta^2}{2}\ +\ \frac{g\,d^2}{2}\ -\ \int_{-d}^{\,\eta}\left[\,\frac{|{\bm\nabla}\phi|^2}{2}\,+\,\frac{\phi_y^{\,2}}{2}\,\right]\mathrm{d}\/y.
\end{equation}
The alternative form (\ref{defL1}) is somehow more convenient. Note that:

\noindent (i) the term $\tilde{\phi}\/\eta_t$, for example, can be replaced by $-\eta\/\tilde{\phi}_t$ after integration by parts;

\noindent (ii) the term $gd^2/2$ can be omitted because, $d$ being prescribed, it does not contribute to the variational principle; 

\noindent (iii) the term $g\eta^2/2$ can be replaced by $gh^2/2$ via a change of definition of $\phi$.

Luke's Lagrangian involves a velocity potential but not explicitly the velocity field. Thus, any approximation derived from (\ref{defL0}) has an irrotational velocity field because the latter is calculated from the relations (\ref{uvphi}). The water wave problem involving several equations, there are {\em a priori} no reasons to enforce  the irrotationality and not, for example, the incompressibility or the surface isobarity or even any combination of these relations. As it is well known in numerical methods, enforcing an exact resolution of as many equations as possible is not always a good idea. Indeed, numerical analysis and scientific computing know many examples when efficient and most used algorithms do exactly the opposite. These so-called {\em relaxation methods\/} have proven to be very efficient for stiff problems. When solving numerically a system of equations, the exact resolution of a few equations does not necessarily ensure that the overall error is reduced: What really matters is that the global error is minimized. A similar idea of relaxation may also apply to analytical approximations, as advocated in  \cite{Clamond2009}.

In order to give us more freedom for building approximations, while keeping an exact formulation, the variational principle is modified (relaxed) by introducing explicitly the horizontal velocity ${\bm u} = {\bm\nabla}\phi$ and the vertical one $v=\phi_y$. The variational formulation can thus be reformulated with the Lagrangian density
\begin{equation}\label{defL2ori}
  \mathscr{L}\ =\ \tilde{\phi}\,\eta_t\ +\ \breve{\phi}\,d_t\ -\ \frac{g\,\eta^2}{2}\ -\, \int_{-d}^{\eta}\left[\frac{{\bm u}^2+v^2}{2}\,+\,{\bm\mu}{\bm\cdot}({\bm\nabla}\phi-{\bm u})\,+\,\nu(\phi_y-v)\right]\mathrm{d}\/y,
\end{equation}
where the Lagrange multipliers $\bm\mu$ and $\nu$ have to be determined. By variations with respect of $\bm u$ and $v$, one finds at once the definition of the Lagrange multipliers: 
\begin{equation}\label{defmunu}
  {\bm\mu}\ =\ {\bm u}, \qquad \nu\ =\ v,
\end{equation}
so $({\bm\mu},\nu)$ is another representation of the velocity field, in addition to $({\bm u},v)$ and $({\bm\nabla}\phi,\phi_y)$. These relations can be substituted into (\ref{defL2ori}), but it is advantageous to keep the most general form of the Lagrangian. Indeed, it allows to choose ansatz for the Lagrange multipliers $\bm\mu$ and $\nu$ that can be different from the velocity field $\bm u$ and $v$. The Lagrangian density (\ref{defL2ori}) involving six dependent variables \{$\eta, \phi, {\bm u}, v, {\bm\mu}, \nu$\} --- while the original Lagrangian (\ref{defL1}) only two ($\eta$ and $\phi$) --- it allows more and different subordinate relations to be fulfilled.

The connection of (\ref{defL2ori}) with the variational formulation of the classical mechanics can be seen applying Green's theorem to (\ref{defL2ori}) that yields another equivalent variational formulation involving the Lagrangian density
\begin{align}\label{defLr}
  \mathscr{L}\ =&\ (\partial_t\/\eta\,+\,\tilde{\bm \mu}\bm{\cdot}\bm{\nabla}\/\eta\,-\,\tilde{\nu})\,\tilde{\phi}\ +\ (\partial_t\/d\,+\,\breve{\bm \mu}\bm{\cdot}\bm{\nabla} d\,+\,\breve{\nu})\, \breve{\phi}\ -\ {\textstyle{1\over2}}\,g\,\eta^2\  \nonumber \\ &+\ \int_{-d}^{\,\eta} \left[\,\bm{\mu}\bm{\cdot}\bm{u} - {\textstyle{1\over2}}\/\bm{u}^2\, + \,\nu\/v\, -\, {\textstyle{1\over2}}\/v^2\, +\,(\bm{\nabla}\bm{\cdot}\bm{\mu} + \partial_y\nu)\,\phi\,\right]\mathrm{d}\/y,
\end{align}
and if the relations (\ref{defmunu}) are used, this Lagrangian density is reduced to
\begin{align}\label{defL4}
  \mathscr{L}\ =&\ (\partial_t\/\eta\,+\,\tilde{\bm u}\bm{\cdot}\bm{\nabla}\/\eta\,-\,\tilde{v})\,\tilde{\phi}\ +\ (\partial_t\/d\,+\,\breve{\bm u}\bm{\cdot}\bm{\nabla} d\,+\,\breve{v})\,\breve{\phi}\ -\ {\textstyle{1\over2}}\,g\,\eta^2\  \nonumber \\ &+\ \int_{-d}^{\,\eta} \left[\,{\textstyle{1\over2}}\/\bm{u}^2\,+\,{\textstyle{1\over2}}\/v^2\, +\,(\bm{\nabla}\bm{\cdot}\bm{u} + \partial_yv)\,
\phi\,\right]\mathrm{d}\/y.
\end{align}
Thus, the classical Hamilton principle is recovered, \ie, the Lagrangian is the kinetic energy minus the potential energy plus constraints for the incompressibility and the boundary impermeabilities.

The Lagrangians (\ref{defL0}), (\ref{defL1}), (\ref{defL2ori}), (\ref{defLr}) and (\ref{defL4}) yield the same exact relations. However, (\ref{defL2ori}), (\ref{defLr}) and (\ref{defL4}) allow the constructions of approximations that are not exactly irrotational, that is not the case (\ref{defL0}) and (\ref{defL1}). This advantage is illustrated below via some simple examples. Further examples can be found in \cite{Clamond2009,Dutykh2011b}.

\section{Examples}

Here, we illustrate the use of the variational principle via some simple examples. For the sake of simplicity, we always consider the pseudo velocities equal to the velocity, \ie, we take $\bm{\mu}=\bm{u}$ and $\nu=v$. We also focus on two-dimensional problems in constant depth, \ie, one horizontal dimension (denoted $x$) with $d>0$ independent of $t$ and $x$. For brevity, the horizontal velocity is denoted $u$.

\subsection{Shallow water: Serre's equations}\label{secsw}

For surface waves propagating in shallow water, it is well known that the velocity fields varies little along the vertical. A reasonable ansatz for the horizontal velocity is thus one such that $u$ is independent of $y$, \ie, one can consider the approximation
\begin{equation}\label{defuse}
  u(x,y,t)\ \approx\ \bar{u}(x,t), 
\end{equation}
meaning that $u$ is assumed close to its depth-averaged value.\footnote{$\bar{u}= {1\over h} \int_{-d}^\eta u\,\mathrm{d} y$.} In order to introduce a suitable ansatz for the vertical velocity, one can assume, for example, that the fluid incompressibility (\ref{eqincexa}) and the bottom impermeability (\ref{eqimpexa}{\it b}) are fulfilled. These choices lead thus to the ansatz
\begin{equation}\label{defvse}
  v(x,y,t)\ \approx\ -\,(y+d)\,\bar{u}_x.
\end{equation} 
Notice that, with this ansatz, the velocity field is not exactly irrotational, \ie
\begin{equation}
  v_x\ - u_y\ \approx\ -\,(y+d)\,\bar{u}_{xx}.
\end{equation}
This does {\em not\/} mean that we are modeling a vortical motion but, instead, that we are modeling a potential flow via a velocity field that is not exactly irrotational. This should not be more surprising than, \eg, using an approximation such that the pressure at the free surface is not exactly zero.

With the ansatz (\ref{defuse})--(\ref{defvse}), the vertical acceleration (with $\mathrm{D}/\mathrm{D}t$ being the temporal derivative following the motion) is
\begin{align}
  \frac{\mathrm{D}\,v}{\mathrm{D}\/t}\ &=\ \frac{\partial\,v}{\partial\/t}\ +\ u\,\frac{\partial\,v}{\partial\/x}\ +\ v\,\frac{\partial\,v}{\partial\/y}\ \approx\  -\,v\,\bar{u}_x\ -\ (y+d)\,\frac{\mathrm{D}\,\bar{u}_x}{\mathrm{D}\/t}\ =\ \gamma\,\frac{y+d}{h}, 
\end{align}
where $\gamma$ is the vertical acceleration at the free surface: 
\begin{align} \label{defgamma}
  \gamma\ \equiv\ \left.\frac{\mathrm{D}\,v}{\mathrm{D}\/t}\right|_{y=\eta}\ \approx\ h\left[\,\bar{u}_x^{\,2}\,-\,\bar{u}_{xt}\,-\,\bar{u}\,\bar{u}_{xx}\,\right].
\end{align}
The kinetic energy per water column $\mathscr{K}$ is similarly easily derived
\begin{equation}
  \frac{\mathscr{K}}{\rho}\ =\ \int_{-d}^\eta\frac{u^2+v^2}{2}\,\mathrm{d}\/y\ \approx\ \frac{h\,\bar{u}^2}{2}\ +\ \frac{h^3\,\bar{u}_x^{\,2}}{6}. 
\end{equation}
The Hamilton principle (\ref{defL4}) --- \ie, kinetic minus potential energies plus constraints for incompressibility and boundary impermeabilities --- yields, for this ansatz and after some elementary algebra, the Lagrangian density 
\begin{equation}
  \mathscr{L}\ =\ {\textstyle{1\over2}}\,h\,\bar{u}^2\ +\ {\textstyle{1\over6}}\,h^3\,\bar{u}_x^{\,2}\ -\ {\textstyle{1\over2}}\,g\,h^2\ +\,\left\{\,h_t\,+\left[\,h\,\bar{u}\,\right]_x\,\right\}\tilde{\phi}. 
\end{equation}
The Euler--Lagrange equations for this functional are
\begin{align}
  \delta\tilde{\phi}:\quad & 0\ =\ h_t\ +\,\left[\,h\,\bar{u}\,\right]_x,\label{dLdphi}\\
  \delta\bar{u}:\quad & 0\ =\ \tilde{\phi}\,h_x\ -\ [\,h\,\tilde{\phi}\,]_x\ -\ {\textstyle{1\over3}}\,[\,h^3\,\bar{u}_x\,]_x\ +\ h\,\bar{u}, \\
  \delta h:\quad & 0\ =\ {\textstyle{1\over2}}\,\bar{u}^2\ -\ g\,h\ +\ {\textstyle{1\over2}}\,h^2\,\bar{u}_x^{\,2}\ -\ \tilde{\phi}_t\ +\ \tilde{\phi}\,\bar{u}_x\ -\ [\,\bar{u}\,\tilde{\phi}\,]_x,
\end{align}
thence
\begin{align}
  \tilde{\phi}_x\ &=\ \bar{u}\ -\ {\textstyle{1\over3}}\,h^{-1}\,[\,h^3\,\bar{u}_x\,]_x,  \label{serphix}\\
  \tilde{\phi}_t\ &=\ {\textstyle{1\over2}}\,h^2\,\bar{u}_x^{\,2}\ -\ {\textstyle{1\over2}}\,\bar{u}^2\ -\ g\,h\ +\ {\textstyle{1\over3}}\,\bar{u}\,h^{-1}\,[\,h^3\,\bar{u}_x\,]_x. \label{serphit}
\end{align}
Differentiation of (\ref{serphit}) with respect of $x$ yields, after some algebra, the equation 
\begin{align}
  \left[\,\bar{u}\,-\,{\textstyle{1\over3}}\,h^{-1}(h^3\/\bar{u}_x)_x\,\right]_t\, +\ \left[\,{\textstyle{1\over2}}\,\bar{u}^2\,+\,g\,h\,-\,{\textstyle{1\over2}}\,h^2\,\bar{u}_x^{\,2}\,-\,{\textstyle{1\over3}}\,\bar{u}\,h^{-1}(h^3\/\bar{u}_x)_x\,\right]_x\ &=\ 0, \label{eqqdmbisse}
\end{align}
that can rewritten in the non-conservative form
\begin{equation}\label{eqqdmsemod}
  \bar{u}_t\ +\ \bar{u}\,\bar{u}_x\ +\ g\,h_x\ +\ {\textstyle{1\over3}}\,h^{-1}\,\partial_x\!\left[\, h^2\,\gamma\,\right]\,=\ 0.
\end{equation}
After multiplication by $h$ and exploiting (\ref{dLdphi}), we also derive the conservative equations 
\begin{align}
  \left[\,h\,\bar{u}\,\right]_t\ +\,\left[\,h\,\bar{u}^2\, + \,{\textstyle{1\over2}}\,g\,h^2\,+\,{\textstyle{1\over3}}\,h^2\,\gamma\,\right]_x\ =\ 0.\label{eqqdmfluxse0}
\end{align}
In summary, we have derived the system of equations
\begin{align} 
  h_t\ +\ \partial_x\!\left[\,h\,\bar{u}\,\right]\, &=\ 0,  \label{eqmassse} \\ 
  \partial_t\!\left[\,h\,\bar{u}\,\right]\, +\ \partial_x\!\left[\,h\,\bar{u}^2\,+\,{\textstyle{1\over2}}\,g\,h^2\,+\,{\textstyle{1\over3}}\,h^2\,\gamma\,\right]\, &=\ 0, \label{eqqdmfluxse} \\
  h\,\bar{u}_x^{\,2}\ -\ h\,\bar{u}_{xt}\ -\ h\,\bar{u}\,\bar{u}_{xx}\ &=\ \gamma,
\end{align}
that are the Serre equations. With the Serre equations, the irrotationality is not exactly satisfied, and thus these equations cannot be derived from Luke's variational principle.

Ass*uming small derivatives (\ie, long waves) but not small amplitudes, 
these equations were first derived by Serre \cite{Serre1953a} via a different route. They were independently rediscovered by Su and Gardner \cite{SG1969}, and again by Green, Laws and Naghdi \cite{Green1974}. These approximations being valid in shallow water without assuming small amplitude waves, they are therefore sometimes called {\em weakly-dispersive fully-nonlinear approximation} \cite{Wu2001a} and are a generalization of the Saint-Venant \cite{Stoker1957,Wehausen1960} and of the Boussinesq equations. The variational derivation above is obvious and straightforward. Further details on the Serre equations concerning their properties and numerical resolutions can be easily found in the literature, \eg, \cite{Dutykh2011a, Li2002, Su1980}.

\subsection{Deep water: generalized Klein--Gordon equations}\label{secdw}

For waves in deep water, measurements show that the velocity field varies nearly exponentially along the vertical \cite{Grue2003, Jensen2007}, even for very large unsteady waves (including breaking waves). Thus, this property is exploited here to derive simple approximations for gravity waves in deep water.

Let $\kappa>0$ be a characteristic wavenumber corresponding, \eg, to the carrier wave of a modulated wave group or to the peak frequency of a JONSWAP spectrum. Following the discussion above, it is natural to seek approximations in the form
\begin{equation}\label{ansinf}
  \{\,\phi\,;\,u\,;\,v\,\}\ \approx\ \{\,\tilde{\phi}\,;\,\tilde{u}\,;\,\tilde{v}\,\}\ \mathrm{e}^{\kappa\/(y-\eta)},
\end{equation}
where $\tilde{\phi}$, $\tilde{u}$ and $\tilde{v}$ are functions of $x$ and $t$ that can be determined using the variational principle (with or  without additional constraints). The ansatz (\ref{ansinf}) is certainly the simplest possible that is consistent with experimental evidences.

The ansatz (\ref{ansinf}) substituted into the Lagrangian density (\ref{defL4}) yields
\begin{equation}\label{defL21}
  2\,\kappa\,\mathscr{L}\ =\ 2\/\kappa\,\tilde{\phi}\,\eta_t\ -\ g\,\kappa\,\eta^2\ +\ {\textstyle{1\over2}}\,\tilde{u}^2\ +\ {\textstyle{1\over2}}\,\tilde{v}^2\ -\ (\tilde{\phi}_x\,-\,\kappa\,\tilde{\phi}\,\eta)\,\tilde{u}\ -\ \kappa\,\tilde{v}\,\tilde{\phi}.
\end{equation}
With (or without) subordinate relations, this Lagrangian gives various equations. We present here only the case without further constraints, thus the Euler--Lagrange equations yield
\begin{eqnarray*}
  \delta\,\tilde{u}\/:\quad&& 0\ =\ \tilde{u}\ -\ \tilde{\phi}_x\ +\ \kappa\,\tilde{\phi}\,\eta_x, \\
  \delta\,\tilde{v}\/:\quad&&  0\ =\ \tilde{v}\ -\ \kappa\,\tilde{\phi}, \\
  \delta\,\tilde{\phi}\/:\quad&& 0\ =\ 2\/\kappa\,\eta_t\ +\ \tilde{u}_x\ -\ \kappa\,\tilde{v}\ +\ \kappa\,\tilde{u}\,\eta_x, \\
  \delta\,\eta\/:\quad&& 0\ =\ 2\,g\,\kappa\,\eta\ +\ 2\,\kappa\,\tilde{\phi}_t\ +\ \kappa\,[\,\tilde{\phi}\,\tilde{u}\,]_x.
\end{eqnarray*}
The two first relations imply that this approximation is exactly irrotational and their use in the last two equations gives
\begin{align} 
  \eta_t\ +\ {\textstyle{1\over2}}\,\kappa^{-1}\,\tilde{\phi}_{xx}\ -\ {\textstyle{1\over2}}\, \kappa\,\tilde{\phi}\ &=\ {\textstyle{1\over2}}\,\tilde{\phi}\left[\,\eta_{xx}\,+\,\kappa\,\eta_x^{\,2}\,\right], \label{surimpL21}\\
  \tilde{\phi}_t\ +\ g\,\eta\ &=\ -\/{\textstyle{1\over2}}\,\left[\,\tilde{\phi}\,\tilde{\phi}_x\,-\,\kappa\,\tilde{\phi}^2\,\eta_x\,\right]_x. \label{surisoL21}
\end{align}
Since equations (\ref{surimpL21})--(\ref{surisoL21}) derive from an irrotational motion, they can also be obtained from Luke's Lagrangian (\ref{defL0}) under the ansatz (\ref{ansinf}). That would not be the case if, for example, we had enforced the incompressibility in the ansatz because, here, that leads to a rotational ansatz (see \cite{Clamond2009}, \S4.3).

To the linear approximation, after elimination of $\tilde{\phi}$, equations (\ref{surimpL21})--(\ref{surisoL21}) yield
\begin{equation}\label{gkglin}
  \eta_{tt}\ -\ (g/2\kappa)\,\eta_{xx}\ +\ (g\/\kappa/2)\,\eta\ =\ 0,
\end{equation}
that is a Klein--Gordon equation. For this reason, equations (\ref{surimpL21}) and (\ref{surisoL21}) are named here {\em generalized Klein--Gordon\/} (gKG). The Klein--Gordon equation is prominent in mathematical physics and appears, \eg, as a relativistic generalization of the Schr\"odinger equation. The Klein--Gordon equation (\ref{gkglin}) admits a special ($2\pi/k$)-periodic traveling wave solution
\begin{equation*}
  \eta\ =\ a\,\cos k\/(x-c\/t), \qquad c^2\ =\ g\left(\,k^2\,+\,\kappa^2\,\right)\left/\left(2\,\kappa\,k^2\right).\right.
\end{equation*}
Therefore, if $k=\kappa$ the exact dispersion relation of linear waves (\ie, $c^2=g/k$) is recovered, as it should be. This means, in particular, that the gKG model is valid for spectra narrow-banded around the wavenumber $\kappa$. Further details and properties of the gKG are given in \cite{Clamond2009} (section 4.2) and in \cite{Dutykh2015}.

\subsection{Arbitrary depth}\label{secfd}

A general ansatz, for waves in finite constant depth and satisfying identically the bottom impermeability, is suggested by the linear theory of water waves:
\begin{align}\label{ansfin}
  \phi\ \approx\ \frac{\cosh\kappa\/Y}{\cosh\kappa\/h}\,\tilde{\phi}(x,t), \qquad 
  u\ \approx\ \frac{\cosh\kappa\/Y}{\cosh\kappa\/h}\,\tilde{u}(x,t), \qquad 
  v\ \approx\ \frac{\sinh\kappa\/Y}{\sinh\kappa\/h}\,\tilde{v}(x,t), 
\end{align}
where $Y=y+d$. The parameter $\kappa>0$ is a characteristic wave number to be made precise {\em a posteriori\/}. This ansatz is uniformly valid for all depths because it yields the shallow water one (\ref{defuse}) as $\kappa\rightarrow0$, and the deep water one (\ref{ansinf}) as $d\rightarrow\infty$. Obviously, the ansatz (\ref{ansfin}) should be valid for wave fields with wavenumber spectra that are narrow-banded around $\kappa$.

Substituting the ansatz (\ref{ansfin}) into (\ref{defL4}), one obtains 
\begin{align}\label{lagfin}
  \mathscr{L}\ =&\ [\,\eta_t\,+\,\tilde{u}\,\eta_x\,]\,\tilde{\phi}\ -\ \frac{g\,\eta^2}{2}\ +\ \frac{\tilde{v}^2}{2}\,\frac{\sinh(2\kappa h)-2\kappa h}{2\kappa\cosh(2\kappa h)-2\kappa}\ +\ \frac{\tilde{\phi}\,\tilde{v}}{2}\left[\,\frac{2\kappa h}{\sinh(2\kappa h)}\,-\,1\,\right] \nonumber \\ &+\,\left[\,\frac{\tilde{u}^2}{2}\,+\,\tilde{\phi}\,\tilde{u}_x\,-\,\kappa\tanh(\kappa h)\,\tilde{\phi}\,\tilde{u}\,\eta_x\,\right]\frac{\sinh(2\kappa h)+2\kappa h}{2\kappa\cosh(2\kappa h)+2\kappa}.
\end{align}
Applying various constraints, one obtains generalized equations including the ones derived in sections \ref{secsw} and \ref{secdw} as limiting cases. In particular, one can derive arbitrary depth generalizations of the Serre and Klein--Gordon equations; these derivations are left to the reader. The main purpose of this section is to illustrate the easiness of deriving approximations uniformly valid for all depths, contrary to perturbation methods with which the two main theories (\ie, Stokes-like and shallow water expansions) have separated validity domains.

\section{Discussion}\label{secdis}

Via simple examples, we have illustrated above the advantage of using a relaxed variational principle. Further examples can be found in \cite{Clamond2009}. The advantages of this approach is greater on variable depth where it is easy to derive simple approximations not derivable from asymptotic expansions \cite{Dutykh2011b}.

Here, we have used the isochoric velocity field (${\bm u},v$) as subordinate condition, but other conditions can be imposed, as well as imposing different conditions on (${\bm u},v$) and (${\bm \mu},\nu$). Indeed, the velocity field (${\bm u},v$) being not more (nor less) physical than the pseudo-velocity field (${\bm \mu},\nu$) and the potential velocity field (${\bm \nabla}\phi,\phi_y$), the constraints can be imposed by combinations of these three fields.

The relaxed variational principle provides a common platform for deriving several approximate equations from the same ansatz in changing only the constraints. Beside the ansatz and the subordinate conditions, no further approximations are needed to derive the equations. Using more general ansatze (\ie, involving more free functions and parameters) and well chosen constraints, one can hopefully derive more accurate approximations.

Although the possibility of using the variational methods without a small-parameter expansion has been overlooked in the context of water waves, it has long been recognized as a powerful tool in Theoretical Physics, in particular in Quantum Mechanics. This approach is even thought in some undergraduate lectures. For instance, from Berkeley's course on Quantum Mechanics 
\cite{Murayama}:
\begin{center}
\begin{minipage}{0.8\textwidth}
{\em -- The perturbation theory is useful when there is a small dimensionlessparameter in the problem, and the system is exactly solvable when the small parameter is sent to zero.} 

{\em -- ... it is not required that the system has a small parameter, nor that the system is exactly solvable in a certain limit. Therefore it has been useful in studying strongly correlated systems, such as the fractional Quantum Hall effect.}  

\end{minipage}
\end{center}
However, in order to be successful, the great power of the variational method needs to be harnessed with skill and care, as it is well-known in Theoretical Physics. Indeed, as quoted in the same lecture on Quantum Mechanics:  
\begin{center}
\begin{minipage}{0.8\textwidth}
{\em -- ... there is no way to judge how close your result is to the true result. The only thing you can do is to try out many Ans\"atze and compare them.} 

{\em -- ... the success of the variational method depends on the initial ``guess'' ... and an excellent physical intuition is required for a successful application.}
\end{minipage}
\end{center}
But it is also well-known that this approach can be very rewarding:
\begin{center}
\begin{minipage}{0.8\textwidth}
{\em --  For example, R. B. Laughlin \cite{Laughlin1983}  proposed a trial wave function that beat other wave functions that had been proposed earlier, such as ``Wigner crystal''.}

{\em -- Once your wave function gives a lower energy than your rival's, you won the race.\footnotemark}
\end{minipage}
\footnotetext{R. B. Laughlin {\em et al.} earned the 1998 Physics Nobel price.}
\end{center} 
Thus, despite its ``dangers'', the variational approach is a tool of choice for modeling water waves, specially for problems when there are no obvious small parameters or if approximations valid for a broad range are needed. We have illustrated these claims in this chapter.


\end{document}